\documentclass[aps,prl,twocolumn,superscriptaddress,10pt]{revtex4-1}
\usepackage{amsmath}
\usepackage{amssymb}
\usepackage{graphicx}
\usepackage{epstopdf}

\begin{document}

\title{Landau-Lifshitz theory of the magnon-drag thermopower}

\author{Benedetta Flebus}
\affiliation{Institute for Theoretical Physics and Center for Extreme Matter and Emergent Phenomena, Utrecht University, Leuvenlaan 4, 3584 CE Utrecht, The Netherlands}
\affiliation{Department of Physics and Astronomy, University of California, Los Angeles, California 90095, USA}
\author{Rembert A. Duine}
\affiliation{Institute for Theoretical Physics and Center for Extreme Matter and Emergent Phenomena, Utrecht University, Leuvenlaan 4, 3584 CE Utrecht, The Netherlands}
\affiliation{Department of Applied Physics, Eindhoven University of Technology, PO Box 513,
5600 MB, Eindhoven, The Netherlands}
\author{Yaroslav Tserkovnyak}
\affiliation{Department of Physics and Astronomy, University of California, Los Angeles, California 90095, USA}

\begin{abstract}
Metallic ferromagnets subjected to a temperature gradient exhibit a magnonic drag of the electric current. We address this problem by solving a stochastic Landau-Lifshitz equation to calculate the magnon-drag thermopower. The long-wavelength magnetic dynamics result in two contributions to the electromotive force acting on electrons: (1) An adiabatic Berry-phase force related to the solid angle subtended by the magnetic precession and (2) a dissipative correction thereof, which is rooted microscopically in the spin-dephasing scattering. The first contribution results in a net force pushing the electrons towards the hot side, while the second contribution drags electrons towards the cold side, i.e., in the direction of the magnonic drift. The ratio between the two forces is proportional to the ratio between the Gilbert damping coefficient $\alpha$ and the coefficient $\beta$ parametrizing the dissipative contribution to the electromotive force.
\end{abstract}

\maketitle

The interest in thermoelectric phenomena in ferromagnetic heterostructures has been recently revived by the discovery of the spin Seebeck effect \cite{uchidaNAT08,*uchidaNATM10}. This effect is now understood to stem from the interplay of the thermally-driven magnonic spin current in the ferromagnet and the (inverse) spin Hall voltage generation in an adjacent normal metal \cite{xiaoPRB10}. Lucassen~\textit{et al.} \cite{lucassenAPL11} subsequently proposed that the thermally-induced magnon flow in a metallic ferromagnet can also produce a detectable (longitudinal) voltage in the bulk itself, due to the spin-transfer mechanism of magnon drag. Specifically, smooth magnetization texture dynamics induce an electromotive force \cite{duinePRB08sp,*tserkovPRB08mt}, whose net average over thermal fluctuations is proportional to the temperature gradient. In this Letter, we develop a Landau-Lifshitz theory for this magnon drag, which generalizes Ref.~\cite{lucassenAPL11} to include a heretofore disregarded Berry-phase contribution. This additional magnon drag can reverse the sign of the thermopower, which can have potential utility for designing scalable thermopiles based on metallic ferromagnets.

Electrons propagating through a smooth dynamic texture of the directional order parameter $\mathbf{n}(\mathbf{r},t)$ [such that $|\mathbf{n}(\mathbf{r},t)|\equiv1$, with the self-consistent spin density given by $\mathbf{s}=s\mathbf{n}$] experience the geometric electromotive force of \cite{duinePRB08sp,*tserkovPRB08mt}
\begin{equation}
F_i = \frac{\hbar}{2} \left( \mathbf{n} \cdot \partial_{t} \mathbf{n} \times \partial_i \mathbf{n} - \beta \partial_{t} \mathbf{n} \cdot \partial_{i} \mathbf{n} \right)
\label{F}
\end{equation}
for spins up along $\mathbf{n}$ and $-F_i$ for spins down. The resultant electric current density is given by
\begin{equation}
j_i= \frac{\sigma_\uparrow-\sigma_\downarrow}{e}\langle F_i\rangle=\frac{\hbar P\sigma}{2e}\langle\mathbf{n} \cdot \partial_{t} \mathbf{n} \times \partial_i \mathbf{n} - \beta \partial_{t} \mathbf{n} \cdot \partial_{i} \mathbf{n}\rangle\,,
\label{jF}
\end{equation}
where $\sigma=\sigma_\uparrow+\sigma_\downarrow$ is the total electrical conductivity, $P=(\sigma_\uparrow-\sigma_\downarrow)/\sigma$ is the conducting spin polarization, and $e$ is the carrier charge (negative for electrons). The averaging $\langle\dots\rangle$ in Eq.~\eqref{jF} is understood to be taken over the steady-state stochastic fluctuations of the magnetic orientation. The latter obeys the stochastic Landau-Lifshitz-Gilbert equation \cite{hoffmanPRB13}
\begin{equation}
s ( 1 + \alpha \mathbf{n} \times ) \partial_{t} \mathbf{n} + \mathbf{n} \times ( H \mathbf{z} + \mathbf{h}) + \sum_i\partial_{i} \mathbf{j}_i=0\,,
\label{llg}
\end{equation}
where $\alpha$ is the dimensionless Gilbert parameter \cite{gilbertIEEEM04}, $H$ parametrizes a magnetic field (and/or axial anisotropy) along the $z$ axis, and $\mathbf{j}_i= - A \mathbf{n} \times \partial_{i} \mathbf{n}$ is the magnetic spin-current density, which is proportional to the exchange stiffness $A$. For $H>0$, the equilibrium orientation is $\mathbf{n}\to-\mathbf{z}$, which we will suppose in the following. The Langevin field stemming from the (local) Gilbert damping is described by the correlator \cite{brownPR63}
\begin{equation}
\langle h_{i} (\mathbf{r}, \omega) h^*_{j} (\mathbf{r}', \omega') \rangle =  \frac{ 2\pi \alpha s\hbar \omega\delta_{ij} \delta (\mathbf{r} - \mathbf{r}') \delta ( \omega - \omega')}{\tanh\frac{\hbar \omega}{2 k_{B} T(\mathbf{r})}}\,,
\label{hh}
\end{equation}
upon Fourier transforming in time: $\mathbf{h}(\omega)=\int dt e^{i\omega t}\mathbf{h}(t)$.

At temperatures much less than the Curie temperature, $T_c$, it suffices to linearize the magnetic dynamics with respect to small-angle fluctuations. To that end, we switch to the complex variable, $n\equiv n_x-in_y$, parametrizing the transverse spin dynamics. Orienting a uniform thermal gradient along the $x$ axis, $T(x)= T + x\partial_x T$, we Fourier transform the Langevin field \eqref{hh} also in real space, with respect to the $y$ and $z$ axes. Linearizing Eq.~\eqref{llg} for small-angle dynamics results in the Helmholtz equation:
\begin{equation}
A ( \partial^{2}_{x} - \kappa^{2}) n(x, \mathbf{q}, \omega) = h(x, \mathbf{q}, \omega), 
\label{He}
\end{equation}
where $\kappa^{2}\equiv q^{2} + [H-(1+i \alpha)s \omega]/A$, $h\equiv h_x-ih_y$, and $\mathbf{q}$ is the two-dimensional wave vector in the $yz$ plane. Solving Eq.~\eqref{He} using the Green's function method, we substitute the resultant $n$ into the expression for the charge current density \eqref{jF}, which can be appropriately rewritten in the following form (for the nonzero $x$ component):
\begin{align}
j_x=&\frac{\hbar P\sigma}{2e}\int\frac{d^{2}\mathbf{q}d \omega}{(2\pi)^3}\omega\nonumber\\
&\times{\rm Re}\frac{(1+i\beta)\langle n(x,\mathbf{q},\omega) \partial_{x} n^{*}(x, \mathbf{q}', \omega')  \rangle}{(2\pi)^3 \delta ( \mathbf{q} - \mathbf{q}') \delta ( \omega - \omega')}\,.
\end{align}
Tedious but straightforward manipulations, using the correlator \eqref{hh}, finally give the following thermoelectric current density:
\begin{equation}
j_x = \frac{ \alpha s  P  \sigma \partial_x T}{4eA^{2} k_{B} T^{2}}
\int \frac{d^{2} \mathbf{q}  d \omega}{(2\pi)^{3}} \frac{(\hbar \omega)^{3}}{\sinh^2\frac{\hbar \omega}{2 k_{B}T}}{\rm Re}\left[(1+i\beta)I\right]\,,
\label{jx}
\end{equation}
where $I(\kappa)\equiv\kappa / |\kappa|^{2}({\rm Re}\,\kappa)^{2}$, having made the convention that ${\rm Re}\,\kappa>0$.

To recast expression \eqref{jx} in terms of magnon modes, we incorporate the integration over $q_x$  by noticing that, in the limit of low damping, $\alpha\to0$,
\begin{equation}
I = \frac{2}{\pi} \int d q_{x} \frac{1 + i q^{2}_{x}/ \alpha \tilde{\omega}}{(\tilde{\omega} - q^{2}_{x} -q^2 - \xi^{-2}  )^{2} + (\alpha \tilde{\omega})^{2}}\,.
\label{I}
\end{equation}
Here, we have introduced the magnetic exchange length $\xi\equiv\sqrt{A/H}$ and defined $\tilde{\omega}\equiv s \omega/ A$. 
After approximating the Lorentzian in Eq.~\eqref{I} with the delta function when $\alpha\ll1$, Eq.~\eqref{jx} can finally be expressed in terms of a dimensionless integral
\begin{equation}
J(a)\equiv\int_{a/\sqrt{2}}^\infty dx \frac{x^{5} \sqrt{2x^{2} - a^{2}}}{\sinh^{2} x^{2}},
\end{equation}
as
\begin{equation}
\mathbf{j}=\left(1-\frac{\beta}{3 \alpha}\right)J\left(\frac{\lambda}{\xi}\right)\frac{k_{B}P\sigma}{\pi^{2}e} \left( \frac{T}{T_{c}} \right)^{3/2} \boldsymbol{\nabla} T\,.
\label{jf}
\end{equation}
Here, $T$ is the ambient temperature, $k_BT_c\equiv A(\hbar/s)^{1/3}$ estimates the Curie temperature, and $\lambda\equiv \sqrt{\hbar A/ s k_{B} T}$ is the thermal de Broglie wavelength in the absence of an applied field. We note that $\alpha,\beta\ll1$ while $\alpha\sim\beta$, in typical transition-metal ferromagnets \cite{tserkovJMMM08}.

For temperatures much larger than the magnon gap (typically of the order of $1$ K in metallic ferromagnets), $\lambda \ll \xi$ and we can approximate $J(\lambda/\xi)\approx J(0)\sim1$. This limit effectively corresponds to the gapless magnon dispersion of $\epsilon_{\mathbf{q}}\equiv\hbar\omega_\mathbf{q}\approx\hbar A q^{2}/s$. Within the Boltzmann phenomenology,  the magnonic heat current induced by a uniform thermal gradient is given by $\mathbf{j}_{Q}= -\boldsymbol{\nabla} T \int [d^{3} \mathbf{q} /(2\pi)^{3}] (\partial_{q_x} \omega_{\mathbf{q}})^{2} \tau(\omega_{\mathbf{q}}) \epsilon_{\mathbf{q}} \partial_T n_{\rm BE}$, where $\tau^{-1}(\omega_{\mathbf{q}})=2 \alpha \omega_{\mathbf{q}}$ is the Gilbert-damping decay rate of magnons (to remain within the consistent LLG phenomenology) and $n_{\rm BE}=[\text{exp}(\epsilon_{\mathbf{q}} / k_{B} T ) - 1 ]^{-1}$ is  the Bose-Einstein distribution function. By noticing that
\begin{equation}
\epsilon_{\mathbf{q}}\partial_T n_{\rm BE}= k_B\left[ \frac{\hbar \omega_{\mathbf{q}} /2k_BT}{\sinh( \hbar \omega_{\mathbf{q}} /2k_BT)} \right]^{2}\,,
\end{equation}
it is easy to recast the second, $\propto\beta$ contribution to Eq.~(\ref{jf}) in the form
\begin{equation}
\mathbf{j}^{(\beta)}=\beta \frac{\hbar P \sigma}{2 e A}\mathbf{j}_{Q}\,,
\end{equation}
which reproduces the main result of Ref.~\cite{lucassenAPL11}.

The magnon-drag thermopower (Seebeck coefficient),
\begin{equation}
S=\left.-\frac{\partial_x V}{\partial_x T}\right|_{j_x=0}\,,
\label{VT}
\end{equation}
corresponds to the voltage gradient $\partial_x V$ induced under the open-circuit condition. We thus get from Eq.~\eqref{jf}:
\begin{equation}
S=\left(\frac{\beta}{3 \alpha}-1\right)J\frac{k_{B}P}{\pi^{2}e} \left( \frac{T}{T_{c}} \right)^{3/2}=\left(\beta-3\alpha\right) \frac{\hbar P\kappa_m}{2 e A}\,,
\label{S}
\end{equation}
where $\kappa_m=(2/3\pi^2)Jk_BA(T/T_c)^{3/2}/\alpha\hbar$ is the magnonic contribution to the heat conductivity. Such magnon-drag thermopower has recently been observed in Fe and Co \cite{watzmanCM16}, with scaling $\propto T^{3/2}$ over a broad temperature range and opposite sign in the two metals. Note that the sign depends on $\beta/\alpha$ and the effective carrier charge $e$.

\begin{figure}[t]
\includegraphics[width=\linewidth]{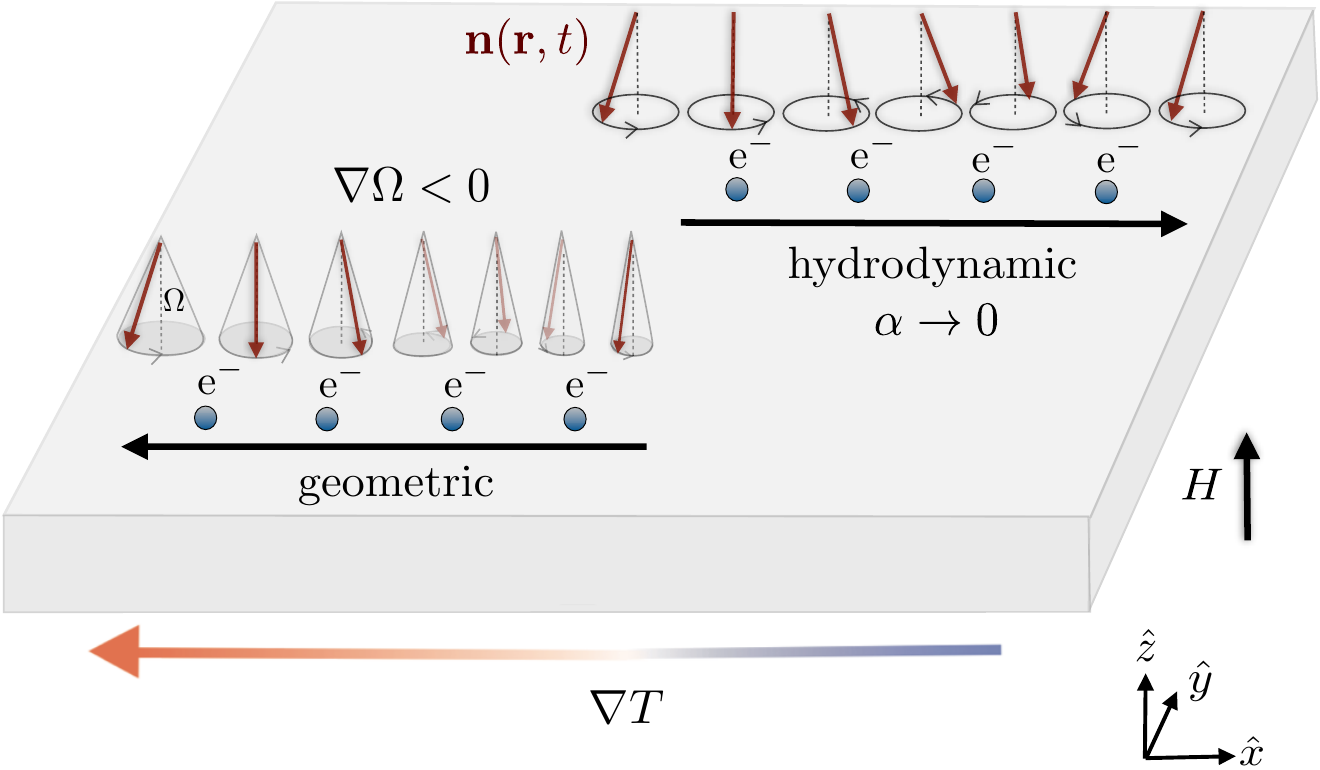}
\caption{Schematics for the two contributions to the electron-magnon drag. In the absence of decay (i.e., $\alpha\to0$), magnons drifting from the hot (left) side to the cold (right) side drag the charge carriers viscously in the same direction, inducing a thermopower $\propto\beta$. The (geometric) Berry-phase drag governed by the magnon decay is proportional to $\alpha$ and acts in the opposite direction. It is illustrated for a spin wave that is thermally emitted from the left. As the spin wave propagates to the right, the solid angle $\Omega$ subtended by the spin precession shrinks, inducing a force oriented to the left for spins parallel to $\mathbf{n}$.}
\label{fig}
\end{figure}

Equations \eqref{jf} and \eqref{S} constitute the main results of this paper. In the absence of Gilbert damping, $\alpha\to0$, the magnon-drag thermopower $S$ is proportional to the heat conductivity. This contribution was studied in Ref.~\cite{lucassenAPL11} and is understood as a viscous hydrodynamic drag. In simple model calculations \cite{tserkovJMMM08}, $\beta P>0$ and this hydrodynamic thermopower thus has the sign of the effective carrier charge $e$. When $P>0$, so that the majority band is polarized along the spin order parameter $\mathbf{n}$, the $\propto\alpha$ contribution to the thermopower is opposite to the $\propto\beta$ contribution. (Note that $\alpha$ is always $>0$, in order to yield the positive dissipation.) The underlying geometric meaning of this result is sketched in Fig.~\ref{fig}. Namely, the spin waves that are generated at the hot end and are propagating towards the cold end are associated with a decreasing solid angle, $\partial_x\Omega<0$. The first term in Eq.~\eqref{F}, which is rooted in the geometric Berry connection \cite{berryPRSLA84,*volovikJPC87,*barnesPRL07,*tserkovPRB09md}, is proportional to the gradient of this solid angle times the precession frequency, $\propto\omega\partial_i\Omega$, resulting in a net force towards the hot side acting on the spins collinear with $\mathbf{n}$.

Note that we have neglected the Onsager-reciprocal backaction of the spin-polarized electron drift on the magnetic dynamics. This is justified as including the corresponding spin-transfer torque in the LLG equation would yield higher-order effects that are beyond our treatment \cite{note}. The diffusive contribution to the Seebeck effect, $\propto T/E_F$, where $E_F$ is a characteristic Fermi energy, which has been omitted from our analysis, is expected to dominate only at very low temperatures \cite{watzmanCM16}. The conventional phonon-drag effects have likewise been disregarded. A systematic study of the relative importance of the magnon and phonon drags is called upon in magnetic metals and semiconductors.

This work is supported by the ARO under Contract No. 911NF-14-1-0016, FAME (an SRC STARnet center sponsored by MARCO and DARPA), the Stichting voor Fundamenteel Onderzoek der Materie (FOM), and the D-ITP consortium, a program of the Netherlands Organization for Scientific Research (NWO) that is funded by the Dutch Ministry of Education, Culture, and Science (OCW).


\begin{thebibliography}{16}
\bibitem [{Note1()}]{note}%
  \BibitemOpen
  \bibinfo {note} {The backaction by the spin-transfer torque would be absent when the longitudinal spin current, $\mathbf{j}_i=\sigma (PE_i+F_i/e)\mathbf{n}$, vanishes, where $E_i$ is the electric field and $F_i$ is the spin-motive force \eqref{F}. Understanding Eq.~\eqref{jf} as pertaining to the limit of the vanishing spin current $\mathbf{j}_i$ rather than electric current $j_i=\sigma (E_i+PF_i/e)\mathbf{n}$ would, however, result in higher-order (in $T/T_c$) corrections to the Seebeck coefficient \eqref{VT}. These are beyond the level of our approximations.}\BibitemShut {Stop}%
\end{thebibliography}
\end{document}